\pdfoutput=1
\documentclass[10pt,twocolumn,a4paper]{IEEEtaes}

\usepackage{color,array,amsthm}
\usepackage{graphicx}

\jvol{XX}
\jnum{XX}
\jmonth{XXXXX}
\paper{1234567}
\pubyear{2022}
\doiinfo{TAES.2022.Doi Number}

\usepackage{cite}
\usepackage{stfloats}
\usepackage{float}
\usepackage{subfigure}
\usepackage{amsmath}
\usepackage{ulem} 
\usepackage{threeparttable}
\usepackage{utfsym}
\usepackage{makecell}
\usepackage{booktabs}
\usepackage{xcolor} 

\usepackage{soul} 
\soulregister{\cite}7 
\soulregister{\ref}7 
\soulregister{\eqref}7 
\soulregister{\usym}7

\hypersetup{ 
	colorlinks=true,
	linkcolor=black,
	filecolor=black,      
	urlcolor=black,
	citecolor=black,
}

\usepackage{tikz}
\usepackage{orcidlink}
\newcommand{\authorcid}[2]{
	\mbox{#1\href{https://orcid.org/#2}{
		\begin{tikzpicture}
			\node at (0,0) []{};
			\node at (0.08,0.095) [inner sep=0pt]
			{\includegraphics[width=0.1in]{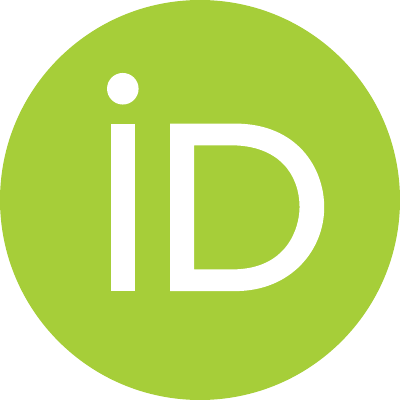}};
		\end{tikzpicture}
		\hspace{-0.08in}
	}}
}

\renewcommand{\refname}{}

\newcommand{\rv}[1]{\textcolor{black}{#1}}

\usepackage{xpatch}
\makeatletter
\ExplSyntaxOn
\cs_new:Npn \bibColoredItems #1#2
{
	\clist_map_inline:nn {#2} { \cs_new:cpn {bib@colored@##1} {#1} } 
}
\ExplSyntaxOff
\newcommand\bib@setcolor[1]{%
	\ifcsname bib@colored@#1\endcsname
	\expanded{\noexpand\color{\csname bib@colored@#1\endcsname}}%
	\else
	\normalcolor
	\fi
}
\IfPackageLoadedTF{hyperref}{\@tempswatrue}{\@tempswafalse}
\if@tempswa
\xpatchcmd\@bibitem {\H@item}{\bib@setcolor{#1}\H@item}{}{\PatchFailed}
\xpatchcmd\@lbibitem{\H@item}{\bib@setcolor{#2}\H@item}{}{\PatchFailed}
\else
\xpatchcmd\@bibitem {\item}  {\bib@setcolor{#1}\item}  {}{\PatchFailed}
\xpatchcmd\@lbibitem{\item}  {\bib@setcolor{#2}\item}  {}{\PatchFailed}
\fi
\makeatother

\begin{document}

\title{RapidPD: Rapid Human and Pet Presence Detection System for Smart Vehicles via Wi-Fi}

\author{\authorcid{HANCHENG GUO}{0009-0005-2213-1604}}
\affil{South China University of Technology, Guangzhou, China} 

\author{\authorcid{ZHEN CHEN}{0000-0001-8018-9103}}
\member{Senior Member, IEEE}
\affil{University of Macau, Macao, China} 

\author{\authorcid{MO HUANG}{0000-0003-0497-194X}}
\member{Senior Member, IEEE}
\affil{University of Macau, Macao, China}

\author{\authorcid{XIUYIN ZHANG}{0000-0003-3659-0586}}
\member{Fellow, IEEE}
\affil{South China University of Technology, Guangzhou, China} 

\receiveddate{Manuscript received XXXXX 00, 0000; revised XXXXX 00, 0000; accepted XXXXX 00, 0000. \\ }

\corresp{{\itshape (Corresponding author: Zhen Chen)}}

\authoraddress{Authors’ address: Hancheng Guo, and Xiuyin Zhang are with School of Electronic and Information Engineering, South China University of Technology, Guangzhou 510006, China (e-mail: \href{mailto:ee_ghch@mail.scut.edu.cn}{ee\_ghch@mail.scut.edu.cn};  \href{mailto:zhangxiuyin@scut.edu.cn}{zhangxiuyin@scut.edu.cn}). Zhen Chen, and Mo Huang are with the State Key Laboratory of Analog and Mixed-Signal VLSI/Institute of Microelectronics,  University of Macau, Macao 999078, China (e-mail: \href{mailto:chenz.scut@gmail.com}{chenz.scut@gmail.com}; \href{mailto:mohuang@um.edu.mo}{mohuang@um.edu.mo}).}


\markboth{GUO ET AL.}{RAPIDPD}
\maketitle

\vspace{0.2in}
\begin{abstract}Heatstroke and \rv{life threatening incidents} resulting from the retention of children and animals in vehicles pose a critical global safety issue. Current presence detection solutions often require specialized hardware or suffer from detection delays that do not meet safety standards. To tackle this issue, \rv{by re-modeling channel state information (CSI) with theoretical analysis of path propagation}, this study introduces RapidPD, an innovative system utilizing \rv{CSI} in subcarrier dimension to detect the presence of humans and pets in vehicles. The system models the impact of motion on CSI and introduces motion statistics in subcarrier dimension using a multi-layer autocorrelation method to quantify environmental changes. RapidPD is implemented using commercial Wi-Fi chipsets and tested in real vehicle environments with data collected from 10 living organisms. Experimental results demonstrate that RapidPD achieves a detection accuracy of 99.05\% and a true positive rate of 99.32\% within a 1-second time window at a low sampling rate of 20 Hz. These findings represent a significant advancement in vehicle safety and provide a foundation for the widespread adoption of presence detection systems.
\end{abstract}
\vspace{0.2in}


\section{INTRODUCTION} \label{sec_intro}

O{\scshape ver} the past decade, device-free passive detection~\cite{youssef2007challenges} has gradually evolved from an emerging technology that allows for the detection of entities without carrying any equipment. To ensure the safety of people's lives and properties, device-free passive detection has been studied and applied in many fields, including intrusion detection, human behavior pattern recognition, and detecting the presence of living organisms in hazardous environments. With the popularity of vehicles, the serious consequences for children or animals due to retention in vehicles have received widespread attention worldwide~\cite{mclaren2005heat, ferrara2013children, costa2016analysis, heatstrokeDeaths, bradley2020environmental, carter2020drugs}. The European New Car Assessment Programme (Euro NCAP) has put forward regulatory requirements for child presence detection (CPD) systems in 2023~\cite{euroncap2025roadmap}. Vehicles equipped with presence detection systems can detect and alert children or pets left alone in the vehicle to avoid heatstroke or even \rv{life threatening incidents}.

Currently, numerous technological solutions are being applied for device-free passive presence detection systems. Early systems for detecting the presence of living organisms were usually based on contact weight or pressure sensors~\cite{rossi2000warning, davis2007child, cole2007system}. Davis~\cite{davis2007child} published a weight sensor-based child presence detection device that is simple to implement but difficult to distinguish inanimate objects from living beings. To address this limitation, capacitive or electrical sensor-based schemes~\cite{george2009seat, ranjan2013child, albesa2014occupancy} have emerged as a more refined solution, for example, Ranjan and George~\cite{ranjan2013child} introduced a child-left-behind warning system based on the capacitive sensing principle. However, this method is constrained by its limited detection range, which is restricted to the seat. In contrast, methods utilizing pyroelectric infrared (PIR) sensors~\cite{mahler2002presence, zappi2010tracking, rashidi2013vehicle} offer a broader detection range through infrared radiation. Despite this advantage, these sensors are prone to temperature fluctuations, which can diminish their reliability in practical applications. To overcome these challenges, Jaworek-Korjakowska \textit{et al.} presented the SafeSO system~\cite{jaworek2021safeso} based on computer vision for seat occupancy classification. Computer vision-based schemes~\cite{cai2017embedded, panda2017multi, fan2016heterogeneous, jaworek2021safeso} are temperature-independent and easily distinguish between living and non-living objects. However, their reliance on specialized camera equipment drives up system costs and raises concerns about potential privacy violations. On the contrary, \rv{radar-based schemes~\cite{abedi2023deep, abedi2021ai, abedi2021passenger, ma2020carosense, innosentRadar, infineonRadar, novelicRadar, ieeRadar, tiRadar}} are valued for their superiority in protecting privacy. \rv{Abedi~\textit{et al.} combined AI with radar technology to achieve in-vehicle occupant detection~\cite{abedi2023deep, abedi2021ai, abedi2021passenger}.} Companies such as InnoSenT~\cite{innosentRadar}, Infineon~\cite{infineonRadar}, NOVELIC~\cite{novelicRadar}, IEE~\cite{ieeRadar}, and Texas Instruments~\cite{tiRadar} have announced their presence detection system-on-chip (SOC) solutions. \rv{The above solutions require additional equipment, ranging from various sensors to millimeter-wave radar. Compared to reusing existing equipment, the additional equipment required to implement a presence detection system increases the cost of the vehicle to varying degrees.}

\rv{Our aim is to investigate the potential of reusing in-vehicle Wi-Fi devices for implementing a presence detection system, which provide a low-cost alternative to traditional solutions. especially in the low-end market accounting for the largest share of the automotive Wi-Fi router market~\cite{wififorecastto2032}. For example, Wi-Fi based solutions~\cite{shi2020no, zeng2022intelligent, zeng2022wicpd, jayaweera2024robust, unimaxwifi} are recognized for their large sensing range and strong privacy protections. Shi~\textit{et al.}~\cite{shi2020no} developed a two-step rear seat child detection system based on commercial Wi-Fi devices using deep learning (DL) method to achieve the distinction between children, pets, and other objects with a detection accuracy of over 95\%.} Zeng \textit{et al.} proposed WiCPD~\cite{zeng2022wicpd}, which introduces a statistical electromagnetic model to explain the effect of motion on all the multipath. \rv{UniMax Electronics Inc~\cite{unimaxwifi} also implemented a CPD system based on Wi-Fi devices. These solution verifies that Wi-Fi is more cost-effective than millimeter-wave radar.}

\rv{Meanwhile, some of the Wi-Fi based solutions explicitly reported their detection latency or window length. \cite{zeng2022intelligent} and WiCPD~\cite{zeng2022wicpd} require a 20-second window to detect a sleeping child, and \cite{unimaxwifi} takes up to a minute to identify whether there is biological movement in the car, neither of which meets Euro NCAP's safety requirement of a 10-second response time. Although \cite{shi2020no} requires only $52\times30$ sized CSI radio images for identification at a transmission rate of $100 pkt/sec$, the sensing range is limited to the rear seat of the vehicle. In addition, some works involving Wi-Fi are also worthy of attention. Li~\textit{et al.} introduced the difference between the CSI solution and the passive radar solution in Wi-Fi sensing technology~\cite{li2020taxonomy}. Tang~\textit{et al.}~\cite{tang2020occupancy}, Li~\textit{et al.}~\cite{li2020passive} and Chen~\textit{et al.}~\cite{chen2016signs} proposed sensing methods with passive Wi-Fi radar. Lyons~\textit{et al.} proposed presence detection in indoor scenarios~\cite{lyons2024wifiact}, which has reference significance for life detection in cars.} Table~\ref{tab_existingStudies}. summarizes the challenges and comparative performance of various presence detection systems across different technologies.

\begin{table*}[!t]
	\caption{Comparison of Existing Works on Presence Detection System \label{tab_existingStudies}}
	\centering
	\tabcolsep = 0.60cm
	\renewcommand\arraystretch{1.1}
	\begin{threeparttable}
		\begin{tabular}{c|c|c|c|c}
			\hline
			Methods & Coverage & \rv{Low-cost\tnote{1}} & Accurate & \rv{Responsive\tnote{2}} \\
			\hline
			Sensors (Weight/Pressure)~\cite{rossi2000warning, davis2007child, cole2007system} & Over Seat	& \usym{2613} & \usym{2613} & Fast \\ 
			\hline
			Sensors (Capacitive/Electrical)~\cite{george2009seat, ranjan2013child, albesa2014occupancy} & Over Seat & \usym{1F5F8} & \usym{2613} & Fast \\
			\hline
			Sensors (PIR)~\cite{mahler2002presence, zappi2010tracking, rashidi2013vehicle} & LoS\tnote{3} & \usym{1F5F8} & \usym{2613} & Moderate \\
			\hline
			Computer Vision (Image/Video)~\cite{cai2017embedded, panda2017multi, fan2016heterogeneous, jaworek2021safeso} & LoS & \usym{2613} & \usym{1F5F8} & Moderate \\
			\hline
			Radar (mmWave)~\cite{abedi2023deep, abedi2021ai, abedi2021passenger, ma2020carosense, innosentRadar, infineonRadar, novelicRadar, ieeRadar, tiRadar} & FoV\tnote{4} & \usym{2613} & \usym{1F5F8} & Fast \\
			\hline
			\rv{A two-step system with DL (Wi-Fi-Based)~\cite{shi2020no}} & \rv{Over Rear Seat} & \rv{$O$} & \rv{\usym{1F5F8}} & \rv{Fast} \\
			\hline
			WiCPD (Wi-Fi-Based)~\cite{zeng2022wicpd} & Whole Car & \rv{$O$} & \usym{1F5F8} & Moderate \\
			\hline
			\rv{UniMax Solution (Wi-Fi-Based)~\cite{unimaxwifi}} & \rv{Whole Car}	& \rv{$O$} & \rv{\usym{1F5F8}} & \rv{Moderate} \\
			\hline
			\textbf{RapidPD} (\rv{\textbf{Ours}}, Wi-Fi-Based) & Whole Car & \rv{$O$} & \usym{1F5F8} & Fast (1sec)\tnote{5} \\
			\hline
		\end{tabular}
		\begin{tablenotes}
			\footnotesize
			\item[1] \rv{Derived from \cite{zeng2022wicpd} and \cite{zeng2022vehicle}, with a \usym{2613} indicating high-cost, a \usym{1F5F8} indicating low-cost, and a $O$ indicating zero-cost reuse of existing equipment.}
			\item[2] The method is referred to as \textit{Fast} if the response is within 10 seconds required by Euro NCAP~\cite{cpdprotocol}, otherwise, it is referred to as \textit{Moderate}.
			\item[3] Line-of-Sight (LoS) \hspace{9em} \tnote{4} File-of-View (FoV) of radar array \hspace{9em} \tnote{5} Length of time window
		\end{tablenotes}
	\end{threeparttable}
	\vspace{-0.06in}
\end{table*}

As mentioned above, presence detection systems deployed in vehicles face additional difficulties due to the complicated multipath and Euro NCAP's requirements for detection delay. We have built the RapidPD system based on commercial Wi-Fi chipsets to achieve rapid presence detection to avoid heatstroke or even \rv{life threatening incidents} in vehicles. In summary, the major contributions of RapidPD are as follows:

\begin{enumerate}
	\def\labelenumi{\arabic{enumi})}
	\item{\rv{
		A CSI model focusing on describing time-varying environments is proposed through a meticulous theoretical analysis of path propagation, which reveals the effect of changing propagation paths on the CSI matrix in subcarrier dimension. The model provides guidance and theoretical basis for utilizing the subcarrier dimension information of CSI.
	}}
	\item{
		\rv{An in-vehicle presence detection system, RapidPD, is developed that uniquely utilizes the subcarrier dimension of CSI. The system introduces a new method for characterizing motion statistics in subcarrier dimension that does not require long windows to accumulate changes, thus extending the range and applicability of Wi-Fi based sensing.}
	}
	\item{
		\rv{The multilayer autocorrelation method is innovatively applied to subcarrier dimension for the proposed RapidPD, which can enhance the detection of weak signals that are masked by the in-vehicle multipath environment.}
	}
	\item{
		The ability of the proposed RapidPD is demonstrated in experiments that can achieve an unprecedented 1-second detection window with over 99.05\% accuracy, offering a valuable solution to prevent heatstroke and \rv{life threatening incidents.}
	}
\end{enumerate}

The remainder of this article is organized as follows. First, the modeling of CSI is introduced in Section~\ref{sec_modeling}. The design of RapidPD is presented in Section~\ref{sec_design} followed by the implementation and evaluation in Section~\ref{sec_evaluation}. Finally, Section~\ref{sec_conclusion} concludes this article.

\section{MODELING OF CSI} \label{sec_modeling}

\begin{figure}[!t]
	\centering
	\includegraphics[width=3.2in]{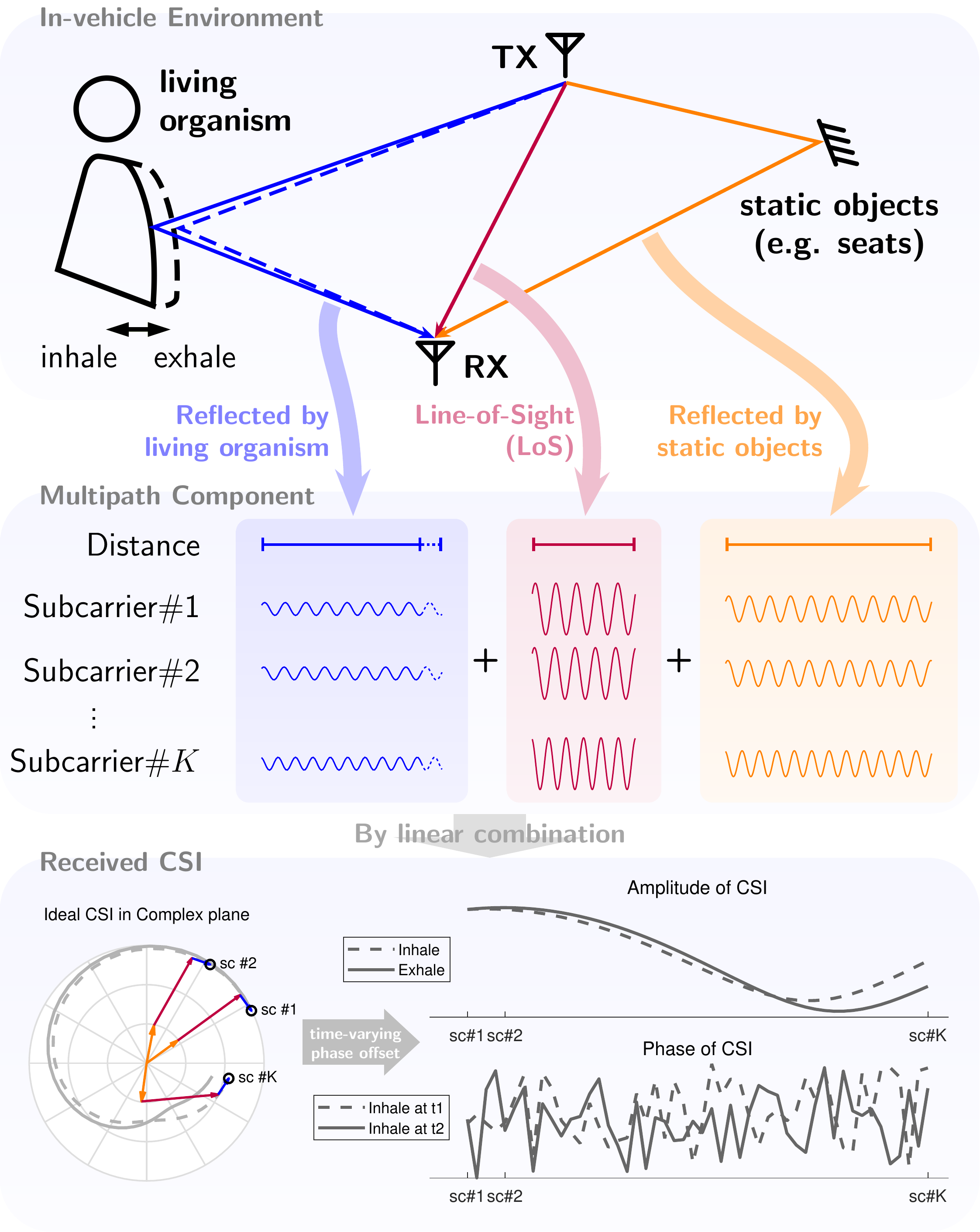}
	\caption{\rv{The effect of living organism's micro-movements on CSI.}}
	\label{fig_abstract}
\end{figure}

Fig.~\ref{fig_abstract} illustrates the effect of micro-movements of living organisms (e.g., breathing while stationary) on CSI. Specifically, subcarriers of different frequencies emitted by the Tx arrive at the Rx via \rv{multiple} paths, each experiencing distinct different amplitude attenuation and phase offsets. Invariant paths correspond to static vectors in the CSI, and \rv{varying} paths result in correlated changes in amplitude and phase across different subcarriers. \rv{Under the influence of the time-varying phase offset~\cite{sen2012you, zhu2018pi, zhou2015wifi, kotaru2015spotfi}, the CSI phase is difficult to utilize because of the instability even for the same state. In contrast, the CSI amplitude preserves the differences caused by linear combinations of CSI vectors corresponding to paths in different states (e.g., inhalation and exhalation), which are correlated in subcarrier direction.}

To develop the modeling of CSI, we first analyze a scenario with a single path featuring only one scatterer, concentrating on how propagation path changes within the environment. \rv{Following this initial analysis, we consider the general case of multiple paths with multiple reflectors, describing how motion manifests itself in subcarrier dimension of the CSI.}

\subsection{The Ideal Static CSI for Commercial Wi-Fi}

Let $X(t,f_i)$ and $Y(t,f_i)$ denote the transmitted and received signals of a subcarrier with frequency $f_i$ at time $t$, where $i \in \Omega_F$ denotes the index of the subcarrier. The estimation equation of CSI $\hat{H}(t,f_i)$ can be expressed as follows~\cite{chiueh2012baseband}:

\begin{equation} \label{eq_hatH}
	\hat{H}(t,f_i) = \rv{\frac{Y(t,f_i)}{X(t,f_i)}},
\end{equation}

\noindent where $X(t,f_i)$ and $Y(t,f_i)$ can be expressed in the form of amplitude and phase, which are $P_X(t,f_i)e^{j\varphi_X(t,f_i)}$ and $P_Y(t,f_i)e^{j\varphi_Y(t,f_i)}$. $P_X(t,f_i)$ and $P_Y(t,f_i)$ denote the power of transmitted and received signals. $\varphi_X(t,f_i)$ and $\varphi_Y(t,f_i)$ denote their phase.

Consider the ideal noise-free static case, where the transmitted signal $X(t,f_i)$ and the received signal $Y(t,f_i)$ are degenerated into $X(f_i)$ and $Y(f_i)$. First, disregarding the effect of noise, assume that there is only one scattering point in the propagation space of the signal. Since the subcarrier frequency interval $\Delta f$ is much smaller than with the Wi-Fi channel center frequency $f_C$ (on the order of GHz), the reflection characteristics of an object can be similar for each subcarrier, with subcarriers of different frequencies experiencing the same path. At this point, according to the radar distance equation~\cite{barton2013radar}, the CSI amplitude $P_X(f_i)$ and $P_Y(f_i)$ extracted from a Wi-Fi device using omnidirectional antennas should be expressed as follows:

\begin{equation} \label{eq_PY}
	P_Y(f_i) = G_{Tx}(f_i) G_{Rx}(f_i) \frac{\sigma_s}{4\pi R_1^2} \frac{\sigma_{Rx}}{4\pi R_2^2} P_X(f_i),
\end{equation}

\noindent where $G_{Tx}(f_i)$ and $G_{Rx}(f_i)$ denote gains obtained from the Tx antenna and Rx antenna respectively for the subcarrier with frequency $f_i$, $\sigma_s$ represents the radar cross-section (RCS) of the scatterer $s$, $\sigma_{Rx}$ represents the effective area of the receiving antenna, and $R_1$ and $R_2$ denote the distances of the transmitting antenna from the scatterer and the scatterer from the receiving antenna.

In addition, the phase of the transmitted signal $\varphi_X(f_i)$ and received signal $\varphi_Y(f_i)$ of a Wi-Fi device using an omnidirectional antenna can be expressed as:

\begin{equation} \label{eq_varphiY}
	\varphi_Y(f_i) = \frac{2\pi f_i}{c} (R_1+R_2) + \pi + \varphi_X(f_i),
\end{equation}

\noindent where $c$ denotes the speed of light, and $\pi$ is half-wave losses during the reflection. Therefore, when there is only one propagation path $l$ and a single scatterer, the estimation equation of CSI $\hat{H}_l(t,f_i)$ can be expressed as follows:

\begin{equation} \label{eq_hatH1p1s} \begin{aligned}
	\hat{H}_l(f_i) = & \rv{\frac{Y(f_i)}{X(f_i)}} \\
	= & G_{Tx}(f_i) G_{Rx}(f_i) \frac{\sigma_s}{4\pi R_1^2} \frac{\sigma_{Rx}}{4\pi R_2^2} \\
	& \exp \left[ j \left( \frac{2\pi f_i }{c} (R_1+R_2) + \pi \right) \right].
\end{aligned} \end{equation}

Considering the general case, there are $M-1$ scatterers along a propagation path $l$, resulting in a total of $M$ propagation segments. Thus, the estimation equation of CSI $\hat{H}_l(t,f_i)$ can be expressed as follows:

\begin{equation} \label{eq_hatH1pss} \begin{aligned}
	\hat{H}_l(f_i) = & G_{Tx}(f_i) G_{Rx}(f_i) \prod_{m=1}^{M} \frac{\sigma_{l,m}}{4\pi R_{l,m}^2} \\
	& \exp \left[ j \left( \frac{2\pi f_i }{c} \sum_{m=1}^{M}R_{l,m} + (M-1)\pi \right) \right],
\end{aligned} \end{equation}

\noindent where $\sigma_{l,m}$ denotes the RCS of scatterer $s_m$ in the propagation path $l$, $\sigma_{l,M}=\sigma_{Rx}$, and $R_{l,m}$ denotes the length of the $m$th segment of propagation path $l$.

Considering the unavoidable multipath situation, the actual CSI is a linear combination of multiple $\hat{H}_l(f_i)$. With $L$ paths in the environment, the estimation equation of CSI $\hat{H}(f_i)$ can be expressed as follows:

\begin{equation} \label{eq_hatHppss}
	\hat{H}(f_i) = \sum_{l=1}^{L}{\hat{H}_l(f_i)}.
\end{equation}

\begin{figure*}[!t]
	\centering
	\includegraphics[width=6.7in]{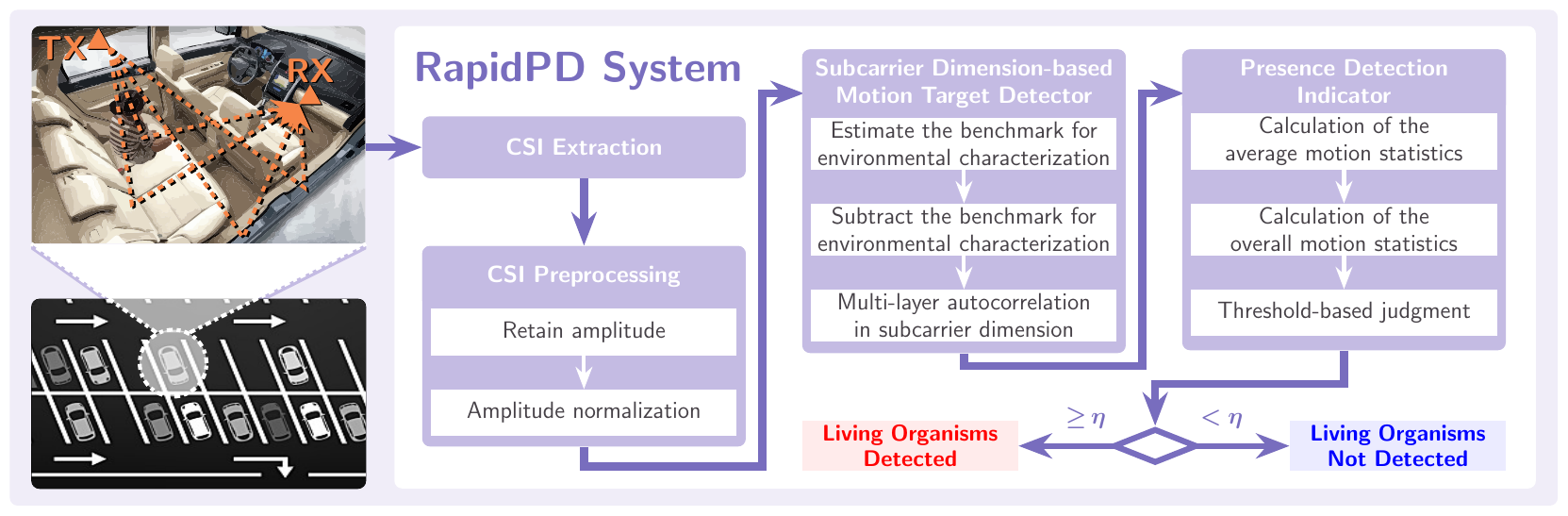}
	\caption{\rv{System architecture of RapidPD.}}
	\label{fig_system}
	\vspace{0.02in}
\end{figure*}

\subsection{Impact of Motion on CSI}

Considering the ideal noise-free dynamic case, we assume that the environment experiences subtle micro-movements over time, which are difficult to detect. For instance, stationary human breathing causes the chest to move slightly, typically between 5 mm and 12 mm~\cite{lowanichkiattikul2016impact}. Given these small movements, the reflections along each signal path are expected to remain mostly unchanged, with only minor variations in the distances the signals travel. Building on this analysis, the estimation equation of time-varying CSI $\hat{H}(t,f_i)$ can be expressed as follows:

\vspace{-0.08in}
\begin{equation} \label{eq_hatHtf} \begin{aligned}
	\hat{H}(t,f_i) =& \sum_{l=1}^{L}{\hat{H}_l(t,f_i)} \\
	= & G_{Tx}(t,f_i) G_{Rx}(t,f_i) \sum_{l=1}^{L} \left\{ \prod_{m=1}^{M} \frac{\sigma_{l,m}}{4\pi R_{l,m}^2(t)} \right. \\
	& \left. \exp \left[ j \left( \frac{2\pi f_i }{c} \sum_{m=1}^{M}R_{l,m}{(t)} + (M-1)\pi \right) \right]  \right\}.
\end{aligned} \end{equation} 

Since the changes in length of propagation paths caused by micro-movements are much shorter than the overall path length in typical application scenarios, the amplitude term in \eqref{eq_hatHtf} can be considered constant over time. Moreover, when we consider the order-of-magnitude relationship between the subcarrier frequency $f_i$, the speed-of-light $c$, and the total distance $\sum_{m=1}^{M}R_{l,m}{(t)}$ of \rv{path $l$}, it becomes evident that these micro-movements are more likely to affect the phase components of \eqref{eq_hatHtf}.

Considering that the gain of Tx and Rx is flat over the channel frequency range, $G_{Tx}(t,f_i)$ and  $G_{Rx}(t,f_i)$ degenerates into $G_{Tx}(t)$ and $G_{Rx}(t)$. Let the variation in the total distance of propagation path $l$ for $t_0$ be $\Delta R_l(t)=\sum_{m=1}^{M} \left[ R_{l,m}(t)-R_{l,m}(t_0) \right]$, and let the invariant term in \eqref{eq_hatHtf} be denoted as $H_l^{\prime}$. The estimation equation of CSI in the ideal noise-free case with micro-movements can be expressed as follows:

\vspace{-0.15in}
\begin{equation} \label{eq_hatHltf}
	\hat{H}(t,f_i) = G_{Tx}(t) G_{Rx}(t) \sum_{l=1}^{L} \rv{H_l^{\prime}(f_i)} \exp \left[ j \frac{2\pi f_i }{c} \Delta R_l(t) \right].
\end{equation} 

From \eqref{eq_hatHltf}, the estimation equation of CSI in the ideal noise-free case with micro-movements can be expressed as a linear combination of complex vectors, each with different phase offsets, after excluding the gains of Tx and Rx. Specifically, the complex vectors \rv{$H_l^{\prime}(f_i)$} correspond to the static environment, while the phase offsets $ \left(2 \pi f_i / c \right) \Delta R_l(t)$ are associated with the time $t$ and subcarrier frequency $f_i$. When the environment changes, the fluctuation in any subcarrier is related to the variation in the total distance $\Delta R_l(t)$ of the propagation path. Additionally, the variation among CSI entries is related to the subcarrier frequency $f$. \rv{It suggests that by finding a benchmark for environmental characterization (e.g., CSI entries at $t_0$),} information on environmental changes can be \rv{sensitively} extracted from the subcarrier dimension \rv{without cumulative change in the time dimension.}


\rv{The above modeling illustrates that environmental changes can lead to correlation changes between different subcarriers of CSI entries. In order to achieve detection of the lifeforms' presence  in the vehicle, a indicator is needed to be constructed to quantify such changes to describe the time-varying environment. The indicator will be covered in the next section on system design. }

\section{RAPIDPD DESIGN} \label{sec_design}


\subsection{System Overview}

Fig.~\ref{fig_system} depicts the overview of RapidPD, which consists of four components including \textit{CSI extraction}, \textit{\rv{CSI preprocessing}}, \textit{subcarrier dimension-based motion target detector}, and \textit{presence detection indicator}. The CSI matrix extracted from the hardware is first \rv{preprocessed to obtain its normalized amplitude.} Afterward, \rv{the benchmark for environmental characterization} is further estimated and subtracted in the motion target detector based on the \rv{normalized} CSI. Subsequently, multi-layer autocorrelation method in subcarrier dimension is applied to the processed CSI entries and the average motion statistics are calculated. Finally, the overall motion statistics are further computed and threshold-based judgments are performed to output a presence detection indication.

\subsection{CSI Preprocessing}

\begin{figure*}[!t]
	\centering
	\subfigure[Unnormalized amplitude on subcarriers with AGC.]{
		\includegraphics[width=3.2in,trim=25 0 30 2,clip]{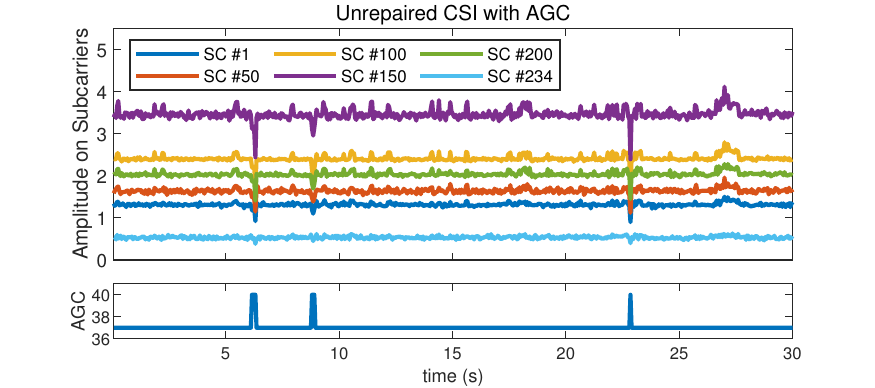}
		\label{fig_amp_t_unrepaired}}
	\hfil
	\hspace{0.1in}
	\subfigure[Normalized amplitude on subcarriers with $s(t)$.]{
		\includegraphics[width=3.2in,trim=25 0 30 2,clip]{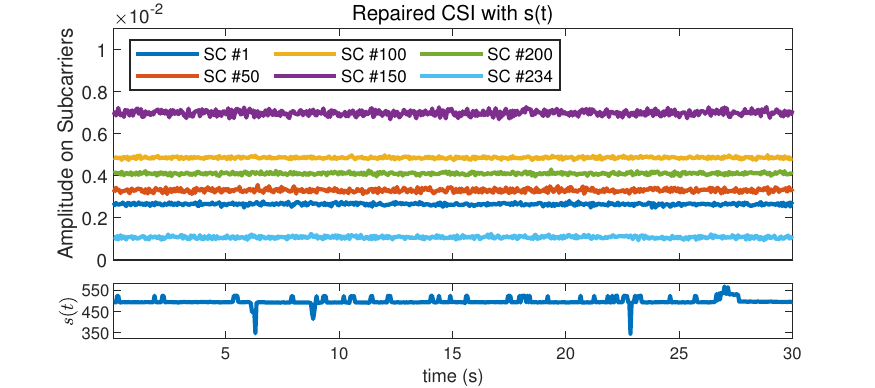}
		\label{fig_amp_t_repaired}}
	\caption{\rv{Examples of CSI amplitude in a static case before and after normalization, with AGC compensation and normalized information $s(t)$.}}
	\label{fig_amp}
	
\end{figure*}

Unlike the ideal case, the measurement of CSI is influenced by unstable perturbations and noise. Based on previous works~\cite{sen2012you, zhu2018pi, zhou2015wifi, kotaru2015spotfi}, the reported unstable perturbations include \rv{time-varying phase offset} and imperfect compensation of automatic gain control (AGC). 

\rv{Considering the time-varying phase offset and the extensive computations involved in complex signals}, we opt to use only the amplitude for detection \rv{referring to existing work~\cite{liu2015tracking, liu2015contactless, zhang2019smars, dahal2024robustness, dahal2024comparison}.} The measurement of CSI amplitude $\left\vert H(t,f_i) \right\vert$ can be expressed as follows:

\begin{equation} \label{eq_Hsimple}
	\rv{
	\left\vert H(t,f_i) \right\vert = G_{agc}(t) \left\vert \hat{H}(t,f_i) \right\vert + \epsilon^{\prime}(t,f_i),
	}
\end{equation} 

\noindent where $\left\vert \cdot \right\vert$ denotes the operations of taking the amplitude taken over the complex signal, $\epsilon^{\prime}(t,f_i)$ is the measurement noise \rv{and $G_{agc}(t)$ denotes the gain of AGC compensation.}

\rv{Due} to the resolution limitations of the hardware, the total gain provided by the low noise amplifier and the programmable gain amplifier \rv{in AGC} cannot fully compensate for the signal's amplitude attenuation. Consequently, the measured amplitude also includes the amplifier's uncertainty error, leading to an amplitude offset.

Amplitude offset is observed in the actual data, even though the amplitude has been compensated by AGC. As shown in Fig.~\ref{fig_amp_t_unrepaired}, notice that the CSI amplitude fluctuates with the AGC field and there are \rv{several significant mutations following a uniform trend across all subcarriers within the window shown}.

Since the CSI amplitude represents the ratio of received to transmitted power, it can be considered received power under the condition that transmitted power is normalized. \rv{As the gain obtained by AGC compensation is the same for each subcarrier,} we can eliminate imperfect compensation by normalizing the total power of the received signal, which is the sum of each CSI entry's amplitude.

First, after obtaining the AGC-compensated amplitudes, the sum of each CSI entry's amplitude $s(t)$ is calculated:

\begin{equation} \label{eq_st}
	s(t) = \sum_{i \in \Omega_F}{\left\vert H(t,f_i) \right\vert}.
\end{equation}

Subsequently, the CSI amplitude was subjected to a normalization operation to obtain the \rv{normalized} amplitude matrix $\tilde{H}(t,f_i)$ as follows:

\begin{equation} \label{eq_tildeH} \begin{aligned}
	\tilde{H}(t,f_i) =& \frac{\left\vert H(t,f_i) \right\vert}{s(t)} \\ 
	=& \rv{\frac{G(t)}{s(t)}\left\vert \sum_{l=1}^{L} H_l^{\prime}(f_i) e^{ j \frac{2\pi f_i }{c} \Delta R_l(t)} \right\vert + \epsilon(t,f_i),}
\end{aligned} \end{equation}

\noindent \rv{where the combined gain $G(t) = G_{agc}(t) G_{Tx}(t) G_{Rx}(t)$ and the noise term $\epsilon(t,f_i) = \epsilon^{\prime}(t,f_i) / s(t)$.}

\rv{As shown in Fig.~\ref{fig_amp_t_repaired}, the normalized amplitude exhibits stability in a static case. Furthermore, as shown in \eqref{eq_tildeH}, the normalized CSI amplitude retains the difference caused by the linear combination of the CSI vectors corresponding to the time-varying paths length $\Delta R_l(t)$, even if the CSI phase is discarded.}

\subsection{Subcarrier Dimension-based Motion Target Detector}

For static environments, the estimation equation of CSI $\hat{H}(t,f_i)$ should be invariant because the electromagnetic wave passes through invariant paths~\cite{zhang2019widetect}. When the unstable Perturbations of CSI are excluded, the variation in the measurement of CSI $H(t,f_i)$ can be attributed to the noise term $\epsilon(t,f_i)$, \rv{as illustrated in Fig.~\ref{fig_amp_t_repaired}.}

The fact that $\epsilon(t,f_i)$ can be approximated as additive Gaussian white noise with zero mean, which is independent both across different times and subcarriers~\cite{zhang2019widetect}, that is, $\epsilon(t_1,f_1)$ and $\epsilon(t_2,f_2)$ are independent for $\forall t_1 \neq t_2, f_1 \neq f_2$.

When there is no detection target in the environment, each CSI entry remains stable, and the variations among CSI entries related to the subcarrier frequency are dominated by the noise $\epsilon(t,f_i)$. Ideally, the noise sequence has no autocorrelation at non-zero lags.

When a detection target with micro-movements is present in the environment, fluctuations in the CSI matrix occur in both the time and subcarrier dimensions. The variation among the CSI entries related to the subcarrier frequency results from both the noise $\epsilon(t,f_i)$ and the variation in the total distance $\Delta R_l(t)$ of the propagation path together. Notably, $\Delta R_l(t)$ in \eqref{eq_tildeH} contributes to a higher autocorrelation at non-zero lags.

To capture these environmental changes and detect the presence of in-vehicle living organisms, we constructively apply the autocorrelation function (ACF) to the CSI in subcarrier dimension. This approach differs from treating each subcarrier as an independent time series~\cite{zeng2022wicpd}, which we refer to as the \textit{time dimension} analysis. Instead, we apply the ACF to each processed CSI entry, which is the \textit{subcarrier dimension} we focus on.

CSI depicts the channel properties of the physical layer in the frequency domain and reveals the combined effects of multipath propagation of the signal, where each CSI entry represents a channel frequency response (CFR)~\cite{chen2020aoa}. For each time window, we average the \rv{preprocessed} CSI matrix in time dimension to smooth the CFR. This averaged CFR is used to \rv{be the benchmark for environmental characterization} $\bar{H}(f_i)$, which can be expressed as follows:

\begin{equation} \label{eq_barH}
	\bar{H}(f_i) = \frac{1}{T} \sum_{t=1}^{T} \rv{\tilde{H}(t,f_i)},
\end{equation}

\noindent where $T$ denotes the length of the time window. 

\begin{figure}[!t] 
	\centering
	\includegraphics[width=3.35in,trim=10 5 20 5]{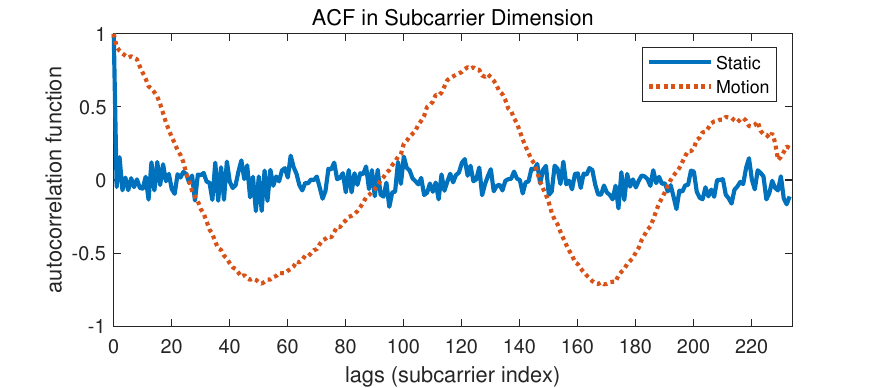}
	\vspace{-0.3cm} 
	\caption{ACF in subcarrier dimension for different cases.}
	\label{fig_acf}
	\vspace{-0.1in}
\end{figure}

In a static environment, \rv{the benchmark for environmental characterization} $\bar{H}(f_i)$ will truthfully characterize the static environment because the noise is smoothed from CFR at this point. In a environment with living organisms, the difference between any CSI entry and $\bar{H}(f_i)$ still has a residual component with high autocorrelation, which we denote as $H_D(t,f_i)$ and expressed as follows:

\begin{equation} \label{eq_HD}
	H_D(t,f_i) = \rv{\tilde{H}(t,f_i)} - \bar{H}(f_i).
\end{equation}

Denote ACF of \rv{$H_D(t,f_i)$} in subcarrier dimension as $\rho(t,\upsilon)$, which is defined as follows:

\begin{equation} \label{eq_rho}
	\rho(t,\upsilon) = \frac{\gamma(t,\upsilon)}{\gamma(t,0)},
\end{equation}

\noindent where $\gamma(t,\upsilon)$ denotes the self-covariance function of the CSI entry at time $t$ as follows:

\begin{equation} \label{eq_gamma}
	\gamma(t,\upsilon) = \mathrm{cov} \rv{\left[ H_D(t,f_i-\upsilon),H_D(t,f_i) \right].}
\end{equation}

In the actual calculation, the sample self-covariance function $\hat{\gamma}(t,\upsilon)$ is used instead of the self-covariance function, which is defined as:

\begin{equation} \label{eq_hatgamma} \begin{aligned}
		\hat{\gamma}(t,\upsilon) = & \hat{\gamma}(t,k \Delta f) \\
		= & \frac{1}{K} \sum_{i=1+k}^{K} \rv{H_D(t,f_{i-k}) H_D(t,f_i),}
\end{aligned} \end{equation}

\noindent where $K$ denotes the total number of subcarriers and let $\upsilon = k \Delta f $.

Using the above method on the actual measured data to calculate $\rho(t,\upsilon)$, the results are shown in Fig.~\ref{fig_acf}. ACF in the static case and dynamic case with human micro-movements are indicated by the blue solid line and the red dashed line. For the static case, the ACF at non-zero lags fluctuates around the zero value. In contrast, for the dynamic cases, the ACF at non-zero lags shows a larger magnitude, making it clearly distinguished from the static scenarios.

Although noise in real systems may not always fully satisfy the independence condition, leading to some autocorrelation at non-zero lags, it is encouraging that the ACF in the subcarrier dimension can still effectively distinguish the presence or absence of living organisms in practical applications. Additionally, we observed that at low levels of Sensing Signal to Noise Ratio (SSNR)~\cite{li2022diversense}, the fluctuations of CSI caused by environmental changes are often masked by noise. To address this, we employ a multi-layer autocorrelation method~\cite{xiaozhi2013inspecting, hou2022weak} to improve the SSNR. Specifically, the signal $\rho_n(t,\upsilon)$ obtained by the $n$-layer ACF in the subcarrier dimension can be expressed as follows:

\vspace{-0.1in}
\begin{equation} \label{eq_rhon}
	\rho_n(t,\upsilon) = \left\{ \begin{array}{lc}
		\dfrac{\rho_{n-1}(t,\upsilon)}{\rho_{n-1}(t,0)} & n \geq 2 \\ \vspace{-0.1in} \\
		\dfrac{\gamma(t,\upsilon)}{\gamma(t,0)} & n = 1
	\end{array} \right. .
\end{equation}

\begin{figure*}[!t]
	\centering
	\subfigure[On additive Gaussian white noise.]{
		\includegraphics[width=3.2in,trim=20 0 25 2,clip]{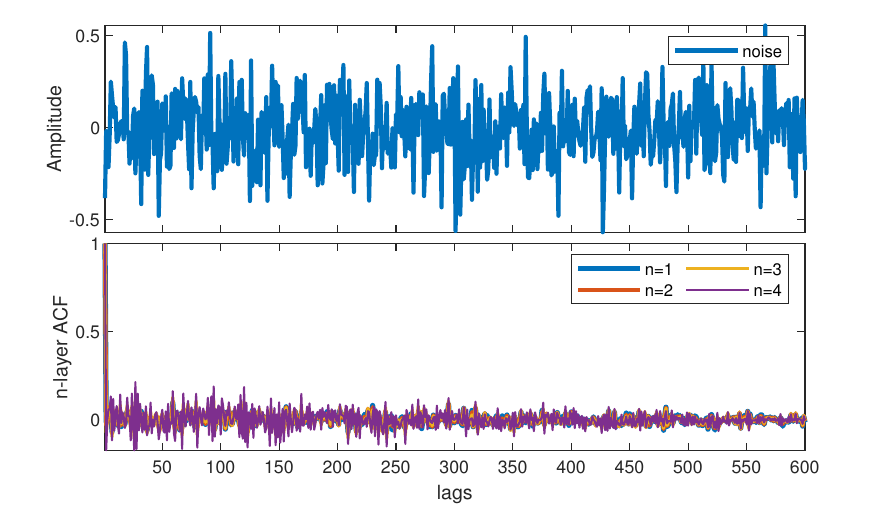}
		\label{fig_multiacf_noise}}
	\hfil
	\hspace{0.1in}
	\subfigure[On noisy sinusoidal signal.]{
		\includegraphics[width=3.2in,trim=20 0 25 2,clip]{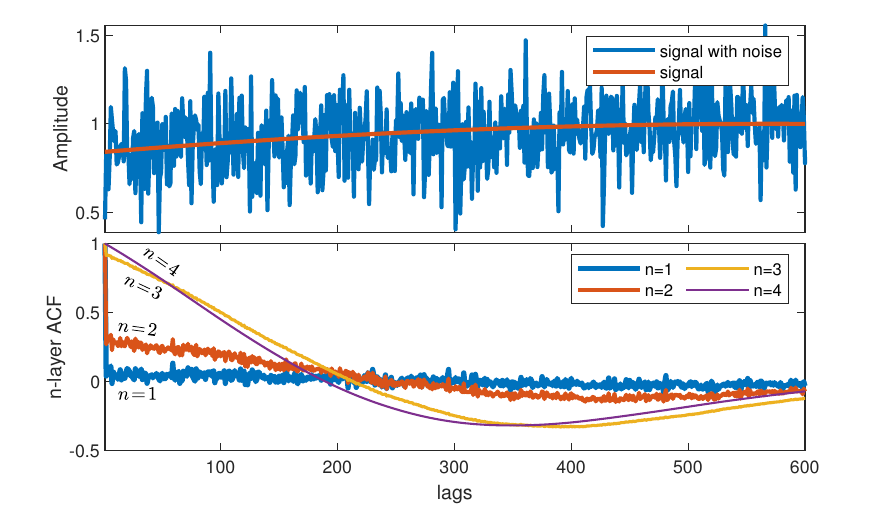}
		\label{fig_multiacf_sin}}
	\caption{\rv{Effectiveness of multi-layer autocorrelation method.}}
	\label{fig_multiacf}
\end{figure*}

\rv{Fig.~\ref{fig_multiacf} illustrates the effectiveness of the multi-layer autocorrelation method in the case of low SSNR. When a low-frequency sinusoidal signal is superimposed with an additive Gaussian white noise, the multi-layer autocorrelation effects of the noise and noisy sinusoidal signals are shown in Fig.~\ref{fig_multiacf_noise} and Fig.~\ref{fig_multiacf_sin}, respectively. As the number of layer $n$ increases, the multi-layer ACF of noisy sinusoidal signal and noise are clearly distinguishable from each other gradually. The multi-layer ACF of the noisy sinusoidal signal gradually deviates from the value of zero, while that of the noise remains near the value of zero at non-zero lags.}

\subsection{Presence Detection Indicator}

Based on \rv{multi-layer ACF}, we propose the motion statistics in subcarrier dimension, which is used to measure the changes in the environment and realize the presence detection of in-vehicle living organisms. The motion statistics $\psi_n(t)$ in subcarrier dimension based on $n$-layer \rv{ACF} can be expressed as follows:

\begin{equation} \label{eq_varphi}
	\psi_n(t) = \rho_n(t,\Delta f).
\end{equation}

\begin{figure}[!t]
	\centering
	\includegraphics[width=3.35in,trim=10 5 20 5]{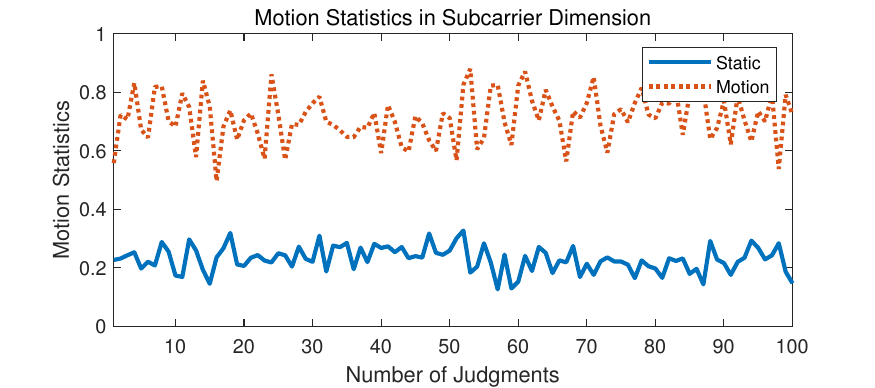}
	\vspace{-0.3cm} 
	\caption{Average motion statistics $\phi$ for different cases.}
	\label{fig_phi}
\end{figure}

Combining all CSI entries in time windows, the average motion statistics $\phi$ in subcarrier dimension on a Tx-Rx stream can be expressed as follows:

\vspace{-0.1in}
\begin{equation} \label{eq_phi}
	\phi = \sum_{t = 1}^{T}{\psi_n(t)}.
\end{equation}

As shown in Fig.~\ref{fig_phi}, we chose $n=3$ to calculate the average motion statistics $\phi$ in subcarrier dimension for both the static case and the dynamic case with human micro-movements. These are indicated by the blue solid line and the red dashed line, respectively. The results show that the average motion statistics $\phi$ in the subcarrier dimension can effectively and clearly distinguish between these two scenarios.

The average motion statistics $\phi$ for the current time window is computed based on a single Tx-Rx stream, and for RapidPD with multiple Tx-Rx streams, $\Phi = \sum \phi$ combines the results of all Tx-Rx streams, which we refer to as the overall motion statistics.

After obtaining the overall motion statistics $\Phi$ for the current time window, a judgment needs to be made based on the set threshold $\eta$. When $\Phi \geq \eta$, an organism is judged to be present, otherwise no organism is.

In practical applications, there may be sudden disturbances that cause data anomalies. Therefore, we obtain the judgments for $m$ windows and take the plural as the final presence detection indication output for smoothing the judgments.

\section{EVALUATION} \label{sec_evaluation}

To comprehensively evaluate RapidPD, we conducted extensive experiments in a typical car and with real infants, children, pets, and adults to validate the detection performance of RapidPD.

\subsection{Methodology}

\begin{figure}[!t]
	\centering
	\includegraphics[height=3.2in, angle=90, viewport=350 1000 1390 3325,clip]{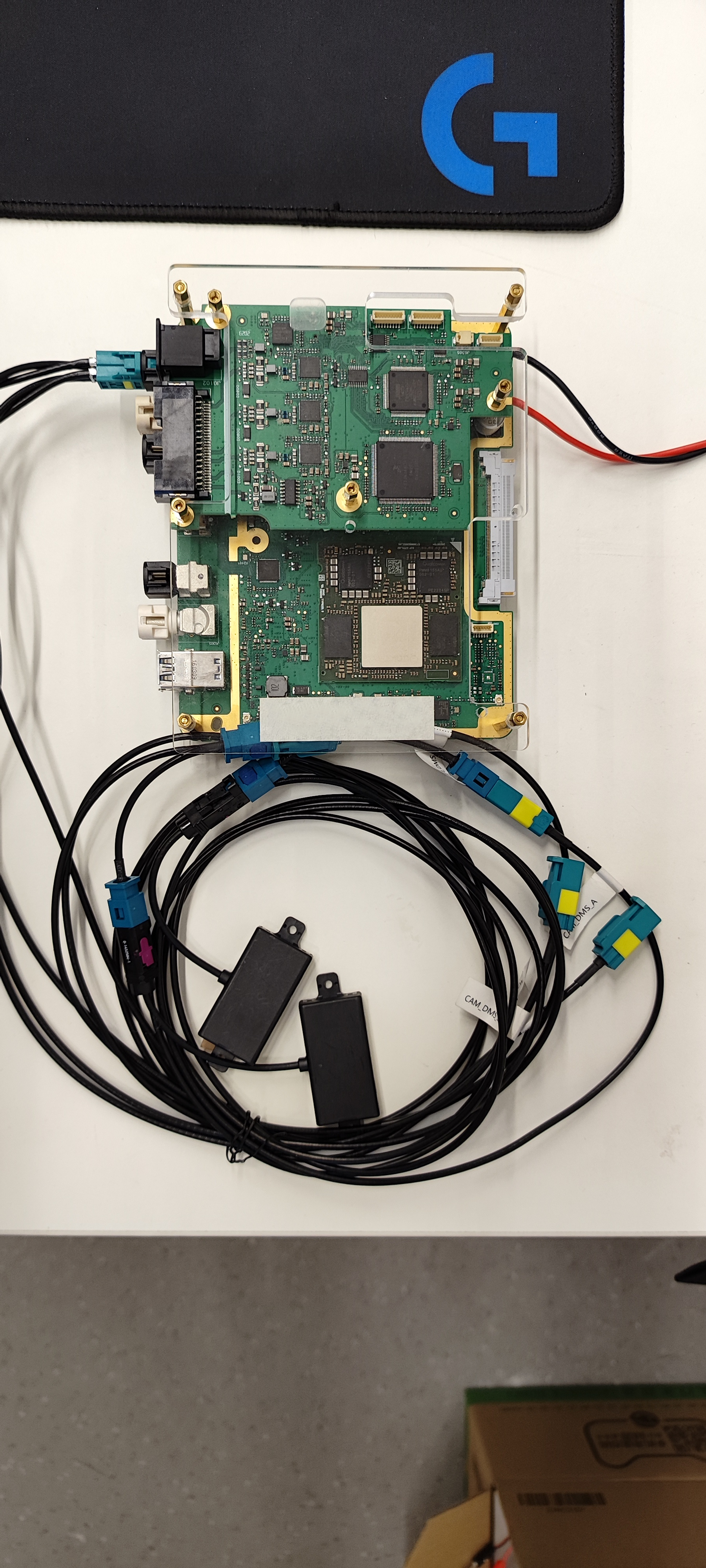}
	\caption{Hardware platform with additional PCB antennas.}
	\label{fig_hardware}
\end{figure}

\begin{figure}[!t]
	\centering
	\subfigure[Tx.]{\includegraphics[width=1.58in,viewport=100 2000 2100 4000,clip]{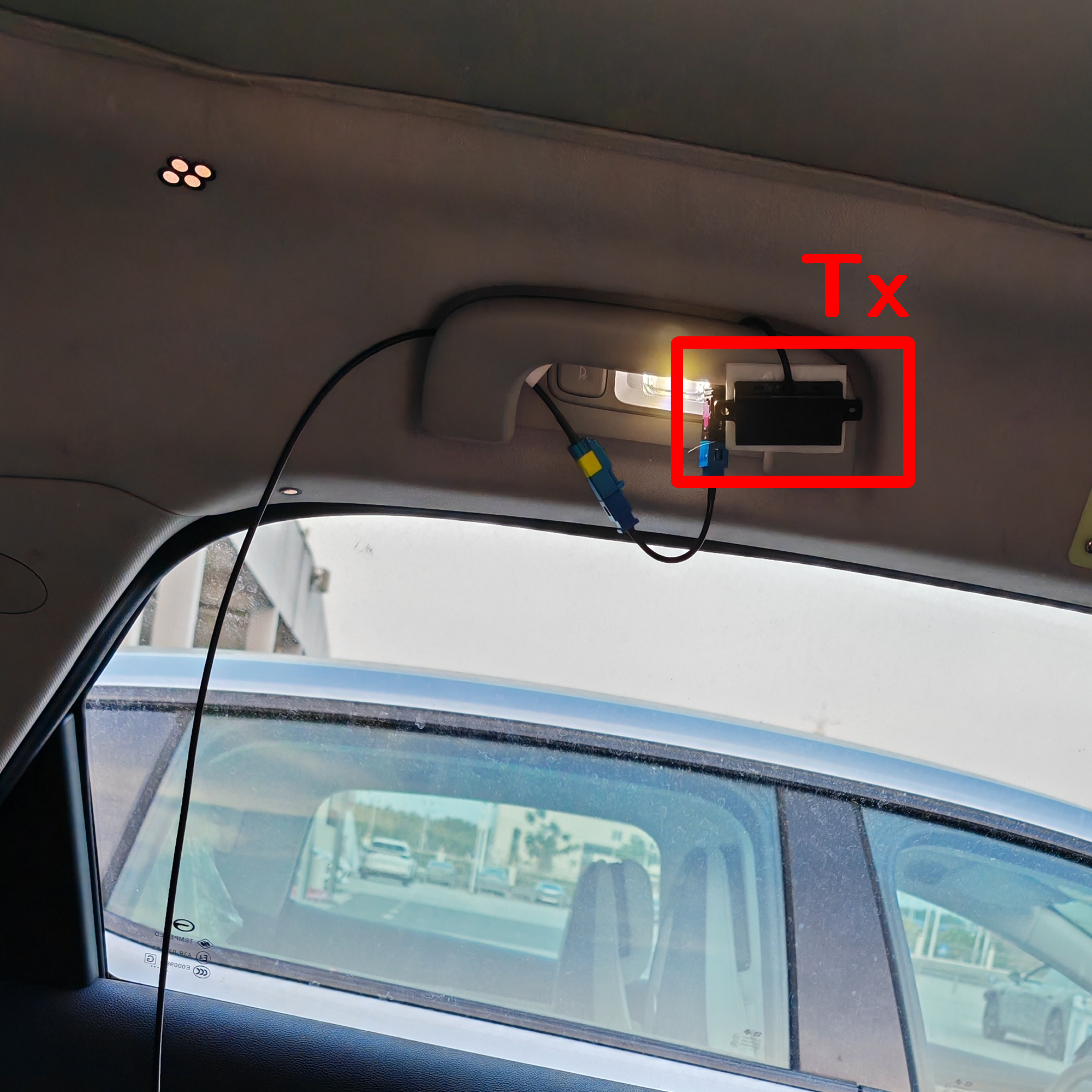}
		\label{fig_tx}}
	\hfil
	\subfigure[Rx.]{\includegraphics[width=1.58in,viewport=400 0 1500 1100,clip]{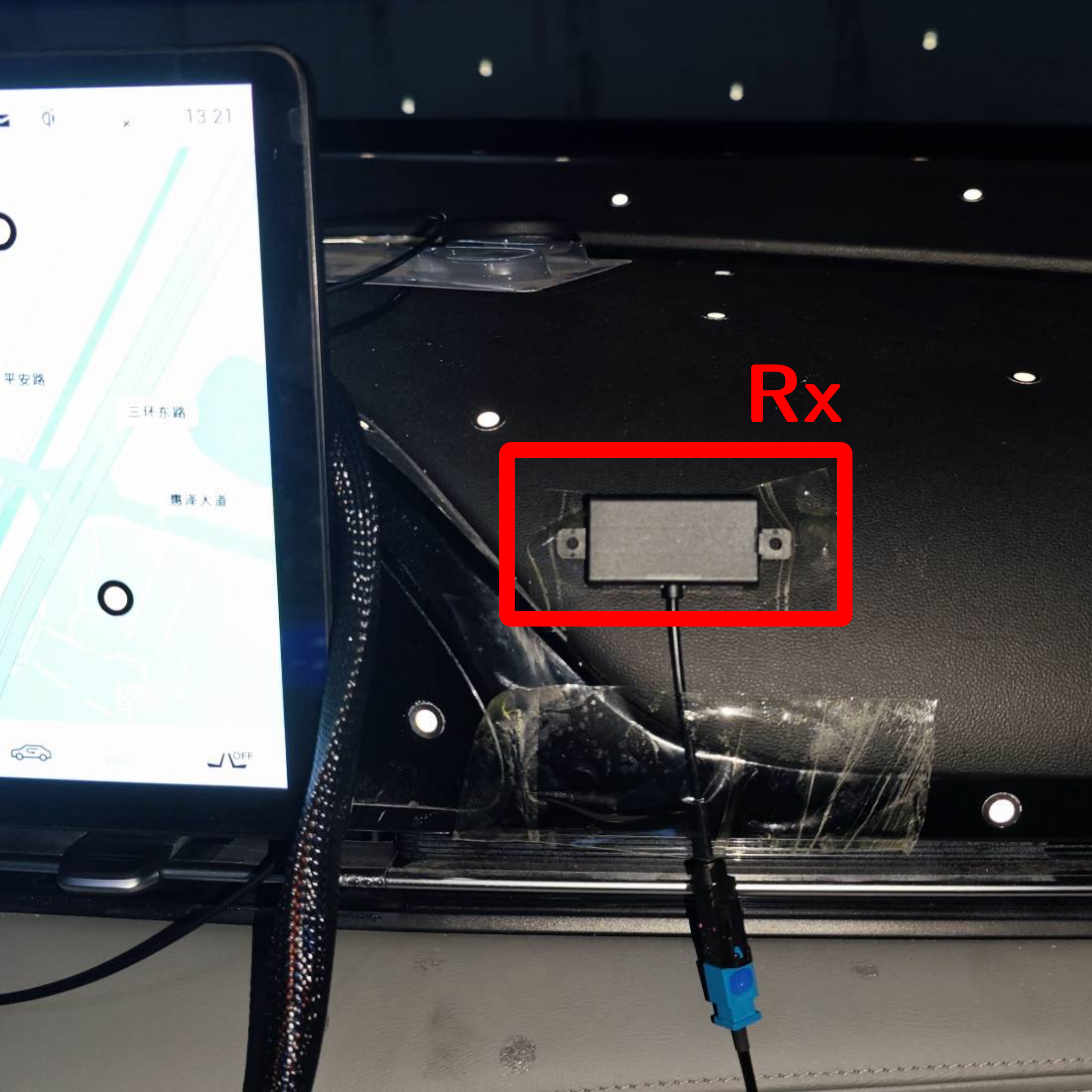}
		\label{fig_rx}}
	\caption{Position of the antenna.}
	\label{fig_antenna}
\end{figure}

\textit{Implementation:} As shown in Fig.~\ref{fig_hardware}, we used a hardware platform based on Infineon's commercial Wi-Fi chipsets CYW8x459 developed by Desay SV with dual bands at 2.4 and 5 GHz and with additional PCB antennas. \rv{RapidPD is deployed on two separate hardware platforms, each carrying a Wi-Fi chip that sends and receives data by programming different customized firmware. An antenna is set up on Rx to receive packets transmitted by two antennas of Tx} at a 20 Hz sampling rate operating on a channel with a center frequency of 5775 MHz (channel 155), which has a bandwidth of 80 MHz and contains 234 obtainable subcarriers. As shown in Fig.~\ref{fig_antenna}, Tx antennas are located at the handles above the rear doors on each side of the vehicle, and the Rx antenna is located on the side of the center console adjacent to the glove box.

RapidPD transfers the data collected in the hardware system to a computer and subsequently processes and analyzes it in MATLAB. To realize an accurate and sensitive presence detection system in vehicles, we take 1s duration data (20 packets at 20Hz sampling rate) as the window and have a 1s window movement step. The number of autocorrelation layers in the motion target detector is chosen as $n=3$, and the number of windows for judgment smoothing in the presence detection indicator is chosen as $m=3$.

\begin{figure}[!t]
	\centering
	\includegraphics[width=3.2in, viewport=350 225 1500 960,clip]{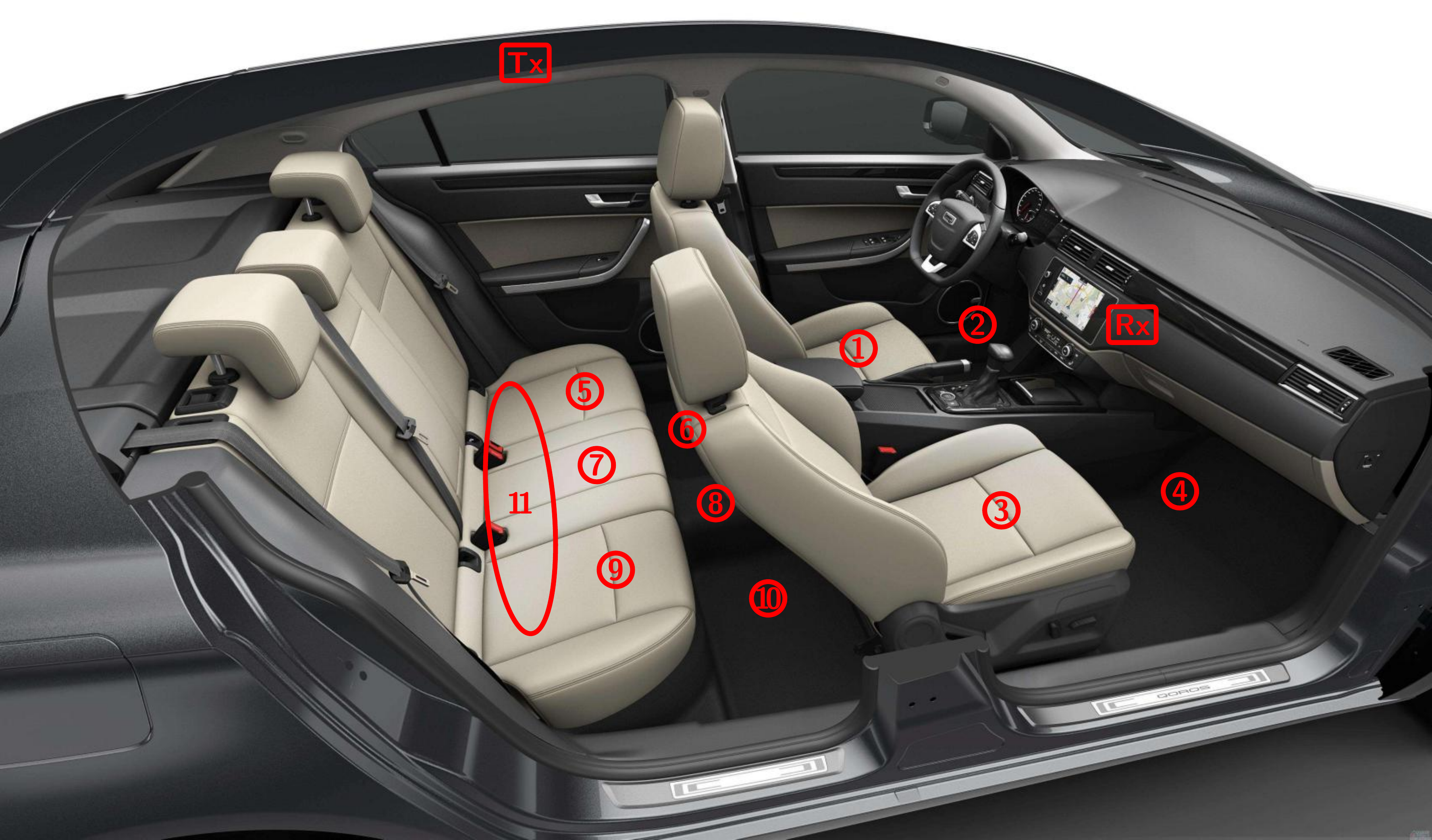}
	\caption{Different test positions for living organisms.}
	\label{fig_position} 
\end{figure}

\begin{figure}[!t]
	\centering
	\subfigure[Dog.]{\includegraphics[width=1.58in,viewport=200 0 1450 824,clip]{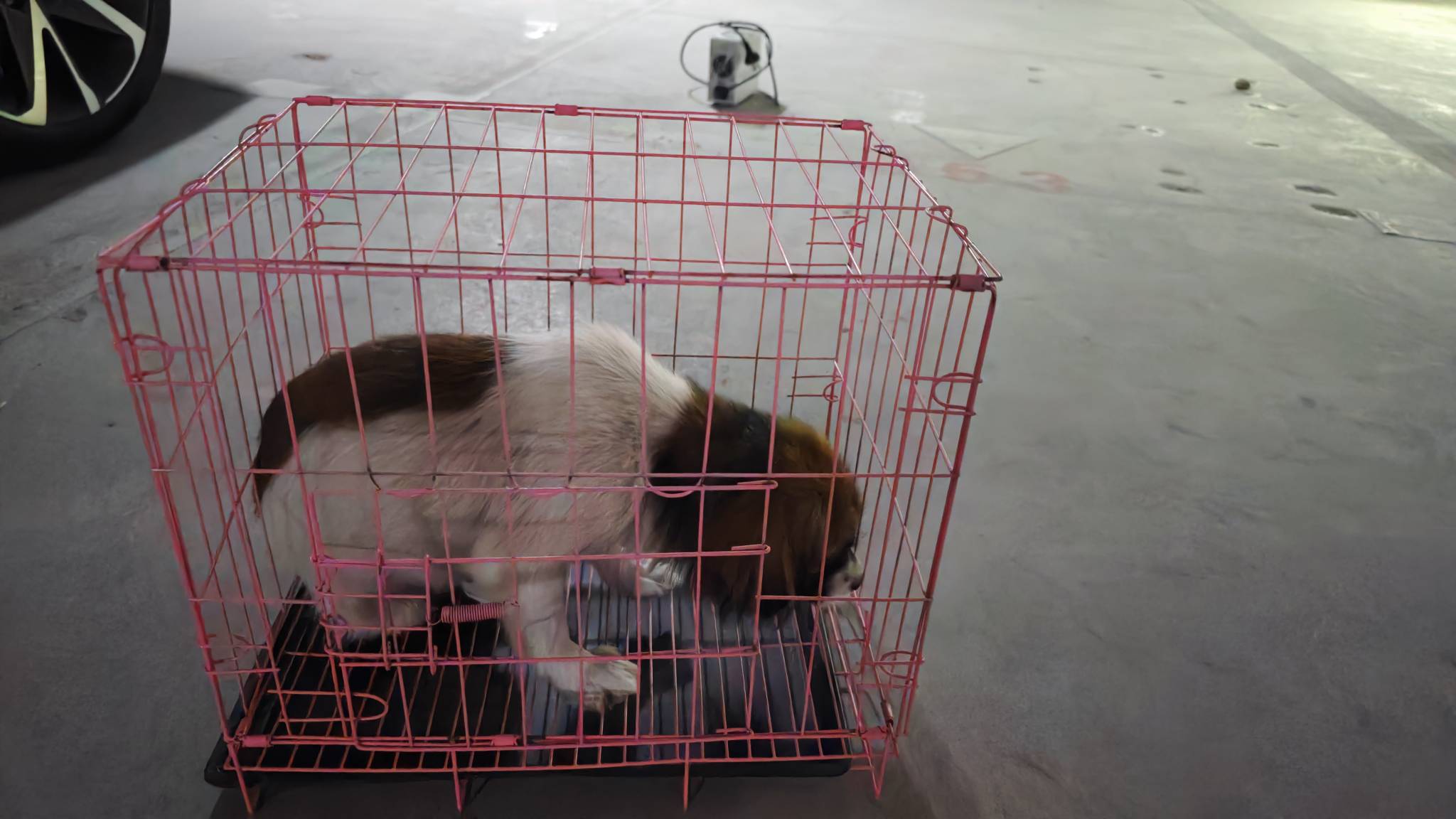}
		\label{fig_dog}}
	\hfil
	\subfigure[Cat.]{\includegraphics[width=1.58in,viewport=450 100 1850 1020,clip]{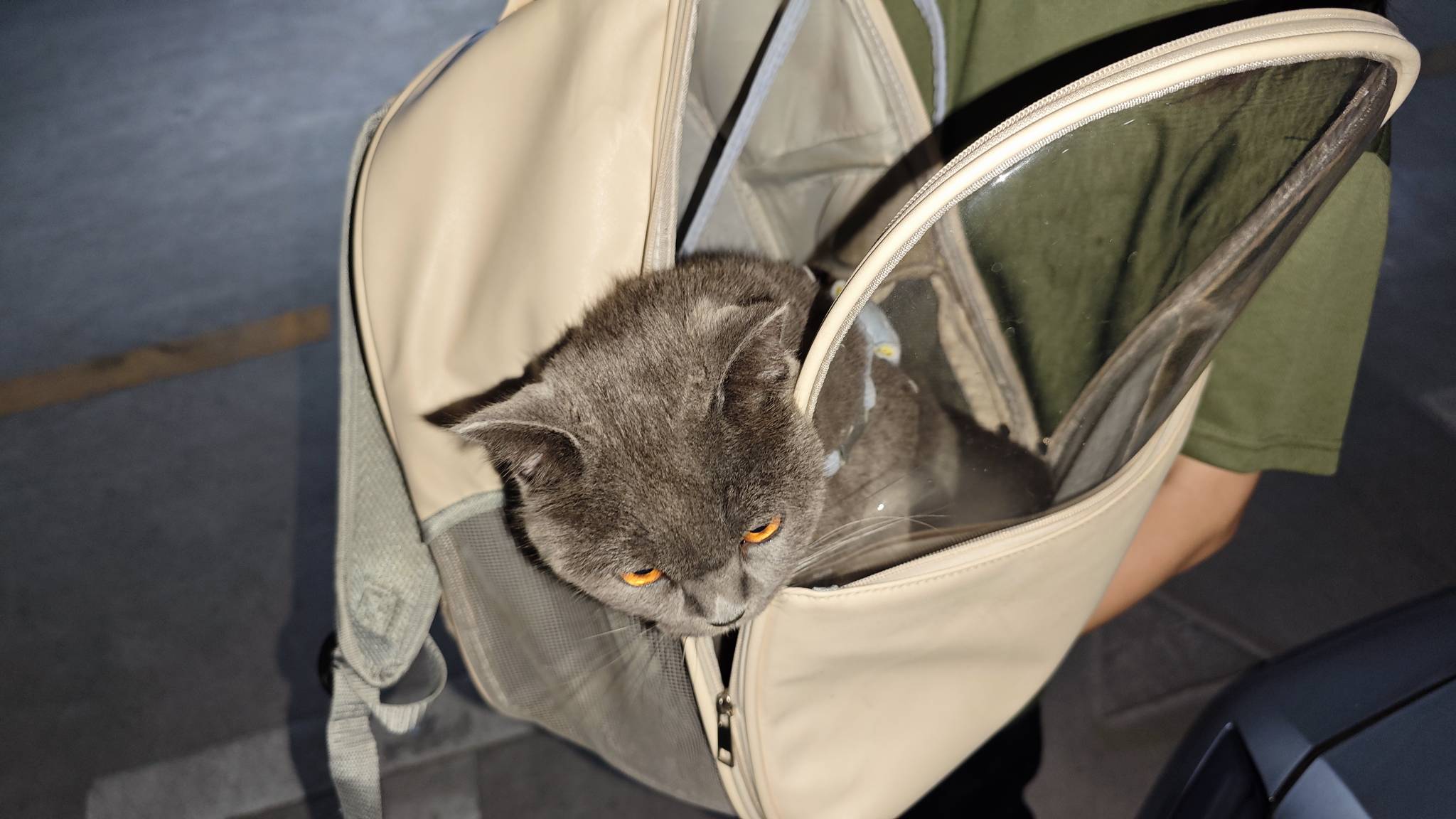}
		\label{fig_cat}}
	\caption{Pets participating in the experiment.}
	\label{fig_pets}
\end{figure}

\textit{Data Collection:} The data collection possessed four main cases including 1)~empty, 2)~human, 3)~dog, and 4)~cat presence. As shown in Fig.~\ref{fig_position}, there are 11 positions in these cases, including 5 seats and corresponding foot positions and rear side seat lie-flat position, which have different types of organisms being tested. The details of the organisms are shown in Table~\ref{tab_lifeformsdetail}, with the pets participating in the experiment shown in Fig.~\ref{fig_pets}.

\begin{table}[!t]
	\caption{The details of organisms \label{tab_lifeformsdetail}}
	\centering
	\tabcolsep = 0.3cm
	\renewcommand\arraystretch{1.1}
	\begin{threeparttable}
		\begin{tabular}{c|c|c|c|c}
			\hline
			\# & Type & Age(Years) & Height(cm) & Weight(kg) \\
			\hline
			1 & Infant & 1 & 74 & 9.5 \\
			\hline
			2 & Child & 3 & 92 & 13.8 \\
			\hline
			3 & Child & 4 & 100 & 14.0 \\
			\hline
			4 & Child & 5 & 114 & 19.0 \\
			\hline
			5 & Child & 5 & 115 & 18.5 \\
			\hline
			6 & Child & 6 & 120 & 26.0 \\
			\hline
			7 & Child & 6 & 120 & 30.0 \\
			\hline
			8 & Adults & - & - & - \\
			\hline
			9 & Dog & - & - & Small-sized \\
			\hline
			10 & Cat & - & - & Medium-sized \\
			\hline
		\end{tabular}
	\end{threeparttable}
\end{table}

\begin{figure*}[!t]
	\centering
	\includegraphics[width=6.7in,trim=30 5 15 10]{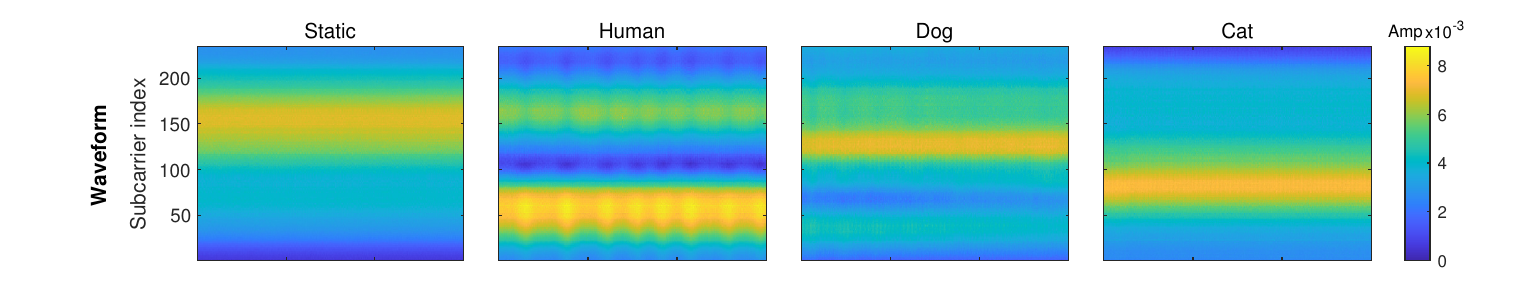}
	\quad
	\includegraphics[width=6.7in,trim=30 5 18 4]{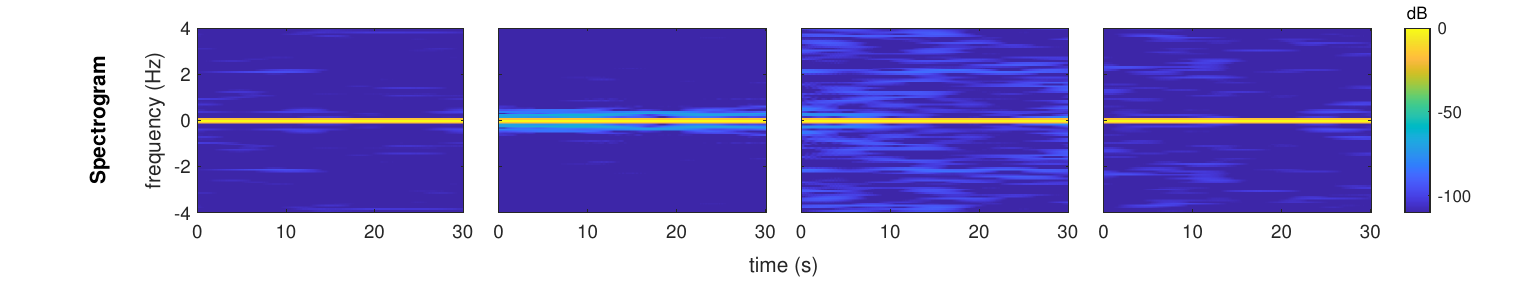}
	\caption{\rv{Sample CSI matrix for each scenario.}}
	\label{fig_csisamp}
\end{figure*}

\begin{figure*}[!t]
	\centering
	\subfigure[Relationship between performance and threshold $\eta$.]
		{\includegraphics[width=3.3in,trim=10 -5 30 5]{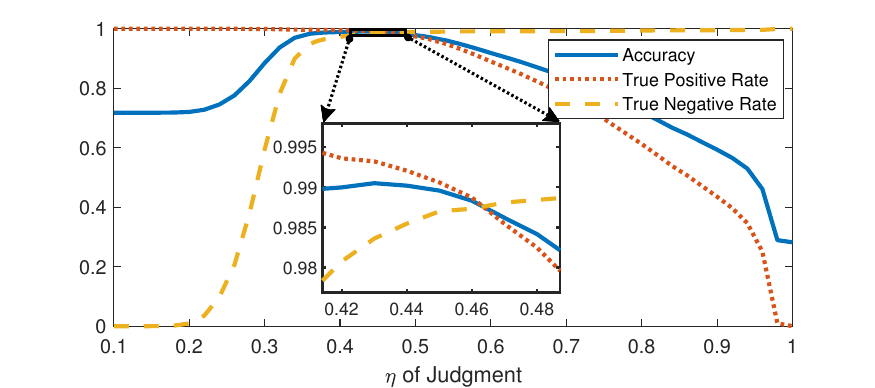}
		\label{fig_res_sum}}
	\hfil
	\subfigure[CDF curve.]
		{\includegraphics[width=3.3in,trim=10 -5 30 5]{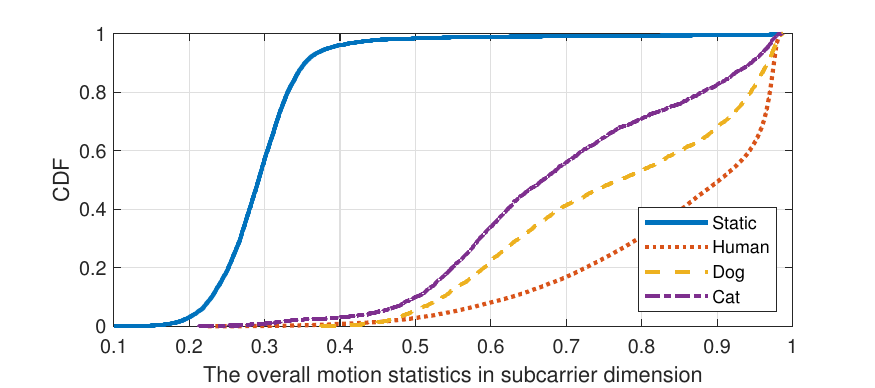}
		\label{fig_res_cdf}}
	\quad
	\vspace{0.1in}
	\subfigure[Judgment accuracy to threshold $\eta$.]
		{\includegraphics[width=3.3in,trim=10 -5 30 5]{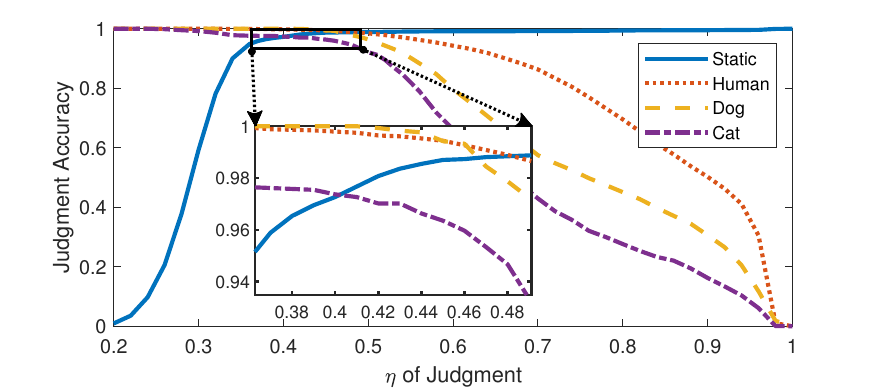}
		\label{fig_res_judgacc}}
	\hfil
	\subfigure[\rv{Confusion matrix for RapidPD judgment at $\eta=0.43$.}]
		{\includegraphics[width=3.3in,trim=-10 -5 -5 0]{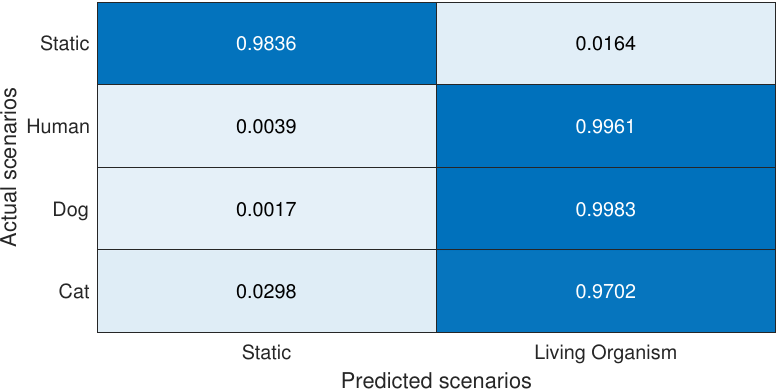}
		\label{fig_res_cm}}
	\caption{Overall performance of RapidPD.}
	\label{fig_res}
\end{figure*}

\rv{Fig.~\ref{fig_csisamp} illustrates sample CSI matrixs for the four scenarios. Spectrograms are generated utilizing STFT with parameters NFFT=256 and OverlapLength=255. The waveforms of the static scenes demonstrate stability, with the human presence scene exhibiting strong respiratory fluctuations and the pet presence scene showing no clearly visible fluctuations. Distinguishably stronger components exist near zero frequency in the spectrogram for human presence scenario, with some cluttered frequency components in other three scenarios, strongest in the dog presence scenario and weakest in the static scenario.}

The experiment was implemented over more than 4 months in different environments, including outdoor open spaces, parking structures, roadsides, and below an elevated bridge. We noted that RapidPD did not need to be altered in the different environments, therefore RapidPD is a calibration-free as well as fast-responding (only 1s of data is needed to complete the judgment) system for human and pet presence detection.

\subsection{Overall Accuracy}

\rv{Fig.~\ref{fig_res} illustrates the overall performance of RapidPD.}

As shown in Fig.~\ref{fig_res_sum}, the accuracy, true positive, and true negative rates vary with the judgment threshold $\eta$. \rv{The accuracy achieved a maximum of 99.05\% at $\eta=0.43$, along with a 99.32\% true positive rate and 1.64\% false positive rate.} Fig.~\ref{fig_res_cdf} illustrates the CDF of the overall motion statistics in the subcarrier dimension of RapidPD, with the living and non-living cases well distinguished. \rv{Fig.~\ref{fig_res_judgacc} shows the curves of the relationship between threshold $\eta$ and the judgment accuracy of the four scenarios. At the selected threshold $\eta=0.43$, all the four cases have high accuracy. Fig.~\ref{fig_res_cm} illustrates the confusion matrix for the judgment case at the chosen threshold, proving that judgment accuracy of the four cases are 98.36\%, 99.61\%, 99.83\%, and 97.02\%.}

The overall accuracy described above was achieved using only a 1-second time window at a low sampling rate of 20Hz, which is an extremely fast response time for a presence detection system and fully meets the Euro NCAP requirement of no more than a 10-second delay.

\subsection{Comparison With Existing Works}

We also implemented a benchmark method that uses a time dimension-based motion target detector~\cite{zhang2019widetect} to replace the subcarrier dimension-based motion target detector proposed in this paper. In addition to the benchmark and RapidPD methods, RapidPD-based methods without multi-layer autocorrelation have been implemented and evaluated as well. The overall accuracy is shown in Table~\ref{tab_compare3methods}. \rv{It is obvious that RapidPD has a great improvement over the benchmark method under the same experimental setup. Compared to removing the multi-layer autocorrelation module, RapidPD obtains a much lower false positive rate.}

\begin{figure}[!t]
	\vspace{0.05in}
	\centering
	\includegraphics[width=3.3in,trim=10 5 30 5]{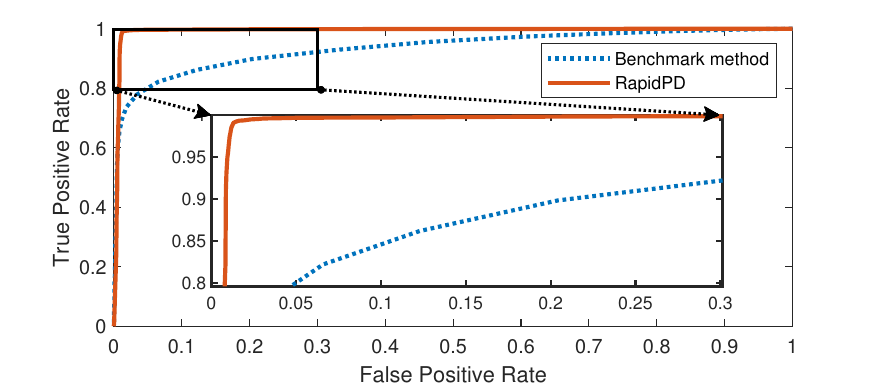}
	\caption{\rv{ROC curves of the benchmark method and RapidPD.}}
	\label{fig_res_roc}
\end{figure}

\begin{table}[!t]
	\caption{Comparing the overall accuracy of motion target detector \label{tab_compare3methods}}
	\centering
	\tabcolsep = 0.28cm
	\renewcommand\arraystretch{1.6}
	\begin{threeparttable}
		\begin{tabular}{c c c c c c}
			\cmidrule(lr){1-6}
			\multicolumn{2}{c}{benchmark method} &
			\multicolumn{2}{c}{\makecell[c]{RapidPD without \\ multi-layer ACF}} &
			\multicolumn{2}{c}{\makecell[c]{\textbf{RapidPD} (with \\ multi-layer ACF)}} \\
			\cmidrule(lr){1-2} \cmidrule(lr){3-4} \cmidrule(lr){5-6}
			\makecell[c]{TPR \\ 89.83\%} & 
			\makecell[c]{FPR \\ 20.41\%} &  
			\makecell[c]{TPR \\ 99.01\%} & 
			\makecell[c]{FPR \\ 3.16\%} & 
			\makecell[c]{TPR \\ \textbf{99.32\%}} & 
			\makecell[c]{FPR \\ \textbf{1.64\%}} \\
			\cmidrule(lr){1-6}
		\end{tabular}
	\end{threeparttable}
	\vspace{-0.1in}
\end{table}
 
Fig.~\ref{fig_res_roc} shows the ROC curves of the benchmark method and RapidPD, \rv{noting that the area under the curve for RapidPD is quite large.} RapidPD possesses a significantly higher true positive rate than the benchmark method with the same false positive rate. This improvement can be attributed to the following reasons:

\begin{enumerate}
	\def\labelenumi{\arabic{enumi})}
	\item{ \rv{
		\textit{Theoretical support derived from re-modeling of CSI:} By analyzing the signal propagation paths, the relevant effects of varying path lengths for different subcarriers are inferred. Compared to accumulating long-term differences in time dimension, the information on environmental changes can be extracted in a shorter time in subcarrier dimension.
	}}
	\item{
		\textit{Combining the information in subcarrier dimension:} RapidPD analyzes the effect of motion on the subcarriers by focusing on their correlation properties rather than examining each subcarrier independently. Each entry in the CSI matrix contains extensive information about the environment, and environmental changes directly impact the correlation between these entries. By leveraging these correlation properties, RapidPD requires only a short time window (1 second) to achieve accurate presence detection.
	}
	\item{
		\textit{Applying the multi-layer autocorrelation method innovatively:} In complex in-vehicle multipath environments, signals undergo multiple reflections before being received, causing motion signals to be more easily drowned out by noise. In cases of low SSNR, RapidPD innovatively applies the multi-layer autocorrelation method, improving accuracy by approximately 0.65\% and reducing the false positive rate by around 1.52\%.
	}
\end{enumerate}

%

\section{CONCLUSION} \label{sec_conclusion}
\rv{By re-modeling CSI with theoretical analysis of path propagation,} this study introduces a novel approach to presence detection leveraging the subcarrier dimensions of the CSI matrix, providing a more precise motion statistics analysis and significantly enhancing detection capabilities. The proposed method based on multi-layer autocorrelation provides a significant indicator for distinguishing the presence or absence of in-vehicle organisms. Extensive experiments validate the effectiveness of RapidPD, demonstrating an accuracy exceeding 99.05\% and a true positive rate greater than 99.32\% using only 1-second time windows at a low-level sampling rate of 20 Hz. This marks the first time subcarrier dimension information from the CSI matrix has been utilized for such sensitive detection, offering a groundbreaking contribution to in-vehicle safety and opening up new possibilities for the global adoption of advanced presence detection systems.

\section*{ACKNOWLEDGMENT}
The authors would like to acknowledge \href{https://en.desaysv.com/}{Desay SV} for providing the hardware equipment and datasets required for this work.

\section*{REFERENCES}
	\vspace{-0.25in}
	\normalem
	\bibliographystyle{IEEEtran}
	\bibColoredItems{black}{
		wififorecastto2032,
		shi2020no,
		jayaweera2024robust,
		dahal2024robustness,
		dahal2024comparison,
		li2020taxonomy,
		tang2020occupancy,
		li2020passive,
		chen2016signs,
		lyons2024wifiact,
		abedi2023deep, 
		abedi2021ai, 
		abedi2021passenger
	}
	\bibliography{IEEEabrv,Reference}

\end{document}